\documentclass[graybox]{svmult}

\usepackage{mathptmx}       
\usepackage{helvet}         
\usepackage{courier}        
\usepackage{type1cm}        
\usepackage{makeidx}         
\usepackage{graphicx}        
\usepackage{multicol}        
\usepackage[bottom]{footmisc}

\makeindex             

\usepackage{amsmath,amsfonts,amssymb}

\usepackage{paralist}
\usepackage{amsbsy}

\usepackage[mathscr]{eucal}

\usepackage{bm}

\usepackage{xspace}

\usepackage{algorithm}
\usepackage{algpseudocode}

\newcommand{\Real}{\mathbb{R}}

\newcommand{\Tra}{^{\sf T}} 
\newcommand{\Inv}{^{-1}} 
\newcommand{\tr}{\operatorname{tr}} 

\newcommand{\V}[1]{{\bm{\mathbf{\MakeLowercase{#1}}}}} 
\newcommand{\VE}[2]{\MakeLowercase{#1}_{#2}} 
\newcommand{\Vn}[2]{\V{#1}^{#2}} 

\newcommand{\M}[1]{{\bm{\mathbf{\MakeUppercase{#1}}}}} 

\newcommand{\Mhat}[1]{{\bm{\hat \mathbf{\MakeUppercase{#1}}}}} 
\newcommand{\Mn}[2]{\M{#1}^{#2}} 

\newcommand{\sign}[1]{\operatorname{sign}({#1})}

\usepackage{amsmath}
\DeclareMathOperator{\y}{\mathbf{y}}
\DeclareMathOperator{\x}{\mathbf{x}}

\DeclareMathOperator{\z}{\mathbf{z}}
\DeclareMathOperator{\uu}{\mathbf{u}}
\DeclareMathOperator{\0}{\mathbf{0}}

\DeclareMathOperator{\X}{\mathbf{X}}
\DeclareMathOperator{\U}{\mathbf{U}}
\DeclareMathOperator{\Z}{\mathbf{Z}}
\DeclareMathOperator{\Ib}{\mathbf{I}}

\DeclareMathOperator{\Y}{\mathbf{Y}}
\DeclareMathOperator{\Hb}{\mathbf{H}}
\DeclareMathOperator{\B}{\mathbf{B}}

\DeclareMathOperator{\bbeta}{\boldsymbol{\beta}}

\usepackage{color, colortbl}
\definecolor{gray}{rgb}{.5,.5,.5}
\definecolor{Tgray}{rgb}{.9,.9,.9}  

\begin{document}

\title*{ADMM Algorithmic Regularization Paths for Sparse
  Statistical Machine Learning} 
\author{Yue Hu, Eric C. Chi and Genevera I. Allen}
\institute{Yue Hu \at Department of Statistics, Rice University, Houston, Texas \\
\email{yh6@rice.edu}
\and Eric C. Chi \at Department of Electrical and Computer Engineering, Rice University, Houston, Texas \\\email{echi@rice.edu}
\and Genevera I. Allen \at Departments of Statistics \& Electrical and
Computer Engineering, Rice University, \& Jan and Dan Duncan
Neurological Research Institute, Baylor College of Medicine and Texas
Children's Hospital, Houston, Texas \\
\email{gallen@rice.edu}
}
%
%
\maketitle

\abstract{
Optimization approaches based on operator splitting are becoming
popular for solving sparsity regularized statistical machine learning
models. While many have proposed fast algorithms to solve these
problems for a single regularization parameter, conspicuously less
attention has been given to computing regularization paths, or 
solving the optimization problems over the full range of
regularization parameters to obtain a sequence of sparse models.   
In this chapter, we aim to quickly approximate the sequence of sparse
models associated with regularization paths for the purposes of
statistical model selection by using the building blocks from a classical
operator splitting method, the Alternating Direction Method of
Multipliers (ADMM).  We begin by proposing an ADMM algorithm
that uses warm-starts to quickly compute the regularization path.
Then, by employing approximations along this warm-starting ADMM algorithm,
we propose a novel concept that we term the {\em ADMM Algorithmic
  Regularization Path}.  Our method can quickly outline the sequence
of sparse models associated with the regularization path in
computational time that is often less than that of using the ADMM
algorithm to solve the problem at a single regularization parameter.
We demonstrate the applicability and substantial computational savings
of our approach through three popular examples, sparse linear
regression, reduced-rank multi-task learning, and convex clustering.  
}

\section{Introduction}
\label{sec:intro}

With the rise of Big Data and the subsequent explosion of
statistical machine 
learning methods to analyze it, statisticians have become avid
consumers of large-scale optimization procedures to estimate sparse
models.  The estimation problem is often cast as an optimization
problem of the form: 
\begin{eqnarray}
\label{eq:the_problem}
\underset{\bbeta}{\text{minimize}}\quad L(\bbeta;\mathbf{W}) + \lambda P(\bbeta),
\end{eqnarray}
where $\bbeta$ is a parameter which specifies a statistical model,
$L(\bbeta;\mathbf{W})$ is a smooth loss function or data-fidelity term
that quantifies 
the discrepancy between the data, $\mathbf{W}$, and the model specified
by $\bbeta$, and $P(\bbeta)$ is a nonsmooth penalty that encourages
sparsity in model parameter $\bbeta$ \cite{boyd2011distributed,buhlmann2011statistics,hastie2009elements}. A regularization parameter,
$\lambda \geq 0$, explicitly trades off the model fit and the 
model complexity.

Directly solving the optimization problem (\ref{eq:the_problem}) is
often challenging. 
Operator splitting methods, such as the Alternating Direction Method
of Multipliers (ADMM), 
have become popular because they convert solving the problem
into solving a sequence of simpler optimization problems that involve only the
smooth loss or nonsmooth penalty.  By breaking up the problem into
smaller ones, ADMM may end up taking more iterations than directly 
solving (\ref{eq:the_problem}), but it often runs in less total time
since the subproblems are typically easy to solve. Clearly in the
context of Big Data, faster algorithms are indispensable, and  
the numerical optimization community has devoted a great deal of
effort to solving (\ref{eq:the_problem}) rapidly for a fixed value of
$\lambda$.  This goal, however, is not necessarily aligned with the
application of statistical machine learning problems to real data.

In practice, statisticians are interested in finding the best sparse
model that represents the data.  Achieving this typically entails a
two-step procedure: (i) model selection, or selecting the best sparse
model or equivalently the best subset
of parameters, and (ii) model fitting, or fitting the model by
minimizing the loss function over the selected parameters \cite{hastie2009elements}.  The first
step is often the most challenging computationally as this entails
searching the combinatorial space of all possible sparse models.  As
this combinatorial search is infeasible for large-scale problems, many
consider convex relaxations through constraints or penalties as
computationally feasible surrogates to help search through the space
of sparse models. Consider for example, sparse linear regression, where
the goal  is to find the subset of variables or inputs that best predicts the
response or output.  Searching over all possible subsets of variables,
however, is an NP hard problem.  Instead, many have employed the
penalty or constraint, $P(\bbeta) = \lVert \bbeta \rVert_{1}$, which is the
tightest convex relaxation to performing best subset selection and whose
solution can be computed in polynomial time.  The nonsmooth penalty
term, $P(\bbeta)$, then serves to translate an infeasible
computational problem into a tractable one for model selection
purposes.

Suppose now that we focus on selecting the best sparse model by means
of penalized statistical estimation as in \eqref{eq:the_problem}. As
$\lambda$ varies, we trace out 
a continuous parametric curve $\hat{\bbeta}(\lambda) \in
\Real^p$. Since this curve cannot 
be determined analytically in general, the curve is estimated for a
finite sequence of regularization parameters. 
To choose the best model, statisticians inspect the sequence of
sparse solutions to 
\eqref{eq:the_problem} over the full 
range of regularization parameters: $\{ \hat{\bbeta}(\lambda_n) : 0
\leq \lambda_1 \leq \cdots \leq \lambda_{\max} \}$, where 
$\lambda_{\max}$ is the value of $\lambda$ at which
$\hat{\bbeta}(\lambda_{\max}) = \mathbf{0}$, the maximally sparse
solution.  This 
sequence of sparse solutions is often called the {\em regularization 
  path} \cite{friedman2007pathwise,friedman2010regularization,hastie2004entire}.
For model selection purposes, however, the actual parameter values,
$\hat{\bbeta}(\lambda)$, as $\lambda$ varies in the regularization
paths are less important than identifying the non-zero components of
$\hat{\bbeta}(\lambda)$.  (Note 
that the parameter values for the optimal model are typically re-fit
anyways in the second model fitting stage.)  
Instead, the support of $\hat{\bbeta}(\lambda)$ or the sequence of
{\em active sets} defined 
as $\mathcal{A}(\lambda) = \{ j : \hat{\beta}_{j}(\lambda) \neq 0
\}$, are the important items; these yield a good sequence of sparse
models to consider that limit computationally intensive exploration of
a combinatorial model space.  Out of this regularization path or
sequence of active sets, the optimal model can be chosen via a number of 
popular techniques such as minimizing the trade-off in model
complexity as with the AIC and BIC, the prediction error as with
cross-validation \cite{hastie2009elements} or the model stability as
with stability 
selection \cite{meinshausen2010stability}.   


To apply sparse statistical learning methods to large-scale
problems, we need fast algorithms not only to fit
\eqref{eq:the_problem} for one value of $\lambda$, but to estimate the
entire sequence of sparse models in the model selection stage.  Our
objective in this chapter is to study the latter, which has received
relatively little attention from the optimization community.
Specifically, we 
seek to develop a new method to approximate the sequence of
active sets associated with regularization paths that is (i) computationally
fast and (ii) comprehensively explores the space of sparse models at a
sufficiently fine resolution. 
In doing so, we will not try to closely approximate the parameter
values, $\hat{\bbeta}(\lambda)$, but instead try to closely
approximate the sparsity of the parameters, $\mathcal{A}(\lambda)$, for
the statistical learning problem \eqref{eq:the_problem}.  

To rapidly approximate the sequence of active sets
associated with regularization paths, we turn to 
the ADMM optimization framework.   We first introduce a procedure to
estimate the regularization path by using the ADMM algorithm with warm  
starts over a range of regularization parameters to yield a path-like
sequence of 
solutions.  Extending this, we preform a one-step approximation along
each point on this path, yielding the novel method that we term
{\em ADMM Algorithmic Regularization Paths}.  Our procedure can
closely approximate active sets given by regularization paths at a
fine resolution, but 
dramatically reduces computational 
time.  
This new approach to estimating a sequence of sparse models opens many
interesting  
questions from both statistical and optimization perspectives.  In
this chapter, however, we focus on motivating our approach and
demonstrating its computational advantages on
several sparse statistical machine learning examples.

This chapter is organized as follows.  We first review how
ADMM algorithms have been used in the statistical machine learning
literature, Section~\ref{sec:review}.  Then, to motivate our approach,
we consider application of ADMM to the familiar example of sparse
linear regression, Section~\ref{sec:dev_path}. 
In Section~\ref{sec:admm_path_general}, we introduce our novel
Algorithmic Regularization 
Paths for general sparse statistical machine learning procedures.  We
then demonstrate how to apply our methods through some popular machine
learning problems in Section~\ref{sec:examples}; specifically, we 
consider 
three examples -- sparse linear regression (Section~\ref{sec:admm_path_lasso}), reduced-rank
multi-task learning (Section~\ref{sec:admm_path_rrr}), and convex clustering
(Section~\ref{sec:admm_path_cvx_clustering}) -- where our Algorithm Paths yield substantial
computational benefits.  We conclude with
a discussion of our work and the many open questions it raises in
Section~\ref{sec:discussion}.

\subsection{ADMM in Statistical Machine Learning}
\label{sec:review}

The ADMM algorithm has become popular in statistical machine learning
in recent 
years because the resulting algorithms are typically simple to code
and can scale efficiently to large problems. 
Although ADMM has been successfully applied over a
diverse spectrum of problems, there are essentially two thematic
challenges among the problems that ADMM has proven adept at
addressing: (i) decoupling constraints and regularizers, that
are straightforward to handle individually, but not in conjunction; and
(ii) simplifying fusion type penalties.  We make note of these two types
of problems because the ADMM Algorithmic Regularization Path we
introduce in this chapter can be applied to either type of problem.

An illustrative example of the first thematic challenge arises in
sparse principal component analysis (PCA). In  \cite{VuChoLei2013} Vu
et al.\@ propose estimating sparse principal subspace estimator
$\Mhat{B}$  
of a symmetric input matrix $\M{S}$ with the solution to the following
semidefinite program:
\begin{eqnarray*}
\underset{\B}{\text{minimize}}\quad - \langle \M{S}, \M{B} \rangle + \lambda \lVert \M{B} \rVert_1 \quad \text{subject to} \quad \M{B} \in \mathcal{F}^d,
\end{eqnarray*}
where $\lVert \M{B} \rVert_1$ is 1-norm of the vectorization of $\M{B}$, the set $\mathcal{F}^d = \{ \M{B} : \M{0} \preceq \M{B} \preceq \M{I}, \tr(\M{B}) = d\}$
is a closed and convex set called the Fantope, and $\lambda \geq 0$ is
a regularization parameter. The main algorithmic challenge is the
interaction between the Fantope constraint and the $\ell_{1}$-norm penalty. If
only either the penalty or constraint were present the problem would
be straightforward to solve. Consider the following equivalent problem
to which ADMM can be readily applied: 
\begin{eqnarray*}
\underset{\B}{\text{minimize}}\quad \delta_{\mathcal{F}^d}(\M{B}) - \langle \M{S}, \M{B} \rangle + \lambda \lVert \M{Z} \rVert_1 \quad \text{subject to}\quad \M{Z} - \M{B} = \M{0},
\end{eqnarray*}
where $\delta_C(\boldsymbol{\Sigma})$ denotes the indicator function
of the closed convex set $C$, namely the function that is $0$ on $C$
and $\infty$ otherwise.  By minimizing an augmented Lagrangian over
$\M{B}$, the copy variable $\M{Z}$, and the scaled dual variable
$\M{U}$ as outlined in \cite{boyd2011distributed}, we arrive at the following
ADMM updates: 
\begin{eqnarray*}
\Mn{B}{k} & = & \underset{\M{B}}{\arg\min}\; \frac{1}{2} \lVert \M{B}
- (\Mn{Z}{k-1} - \Mn{U}{k-1} + \rho\Inv\M{S}) \rVert_{\text{F}}^2 +
\delta_{\mathcal{F}^d}(\M{B})  =  \mathcal{P}_{\mathcal{F}^d}
(\Mn{Z}{k-1} - \Mn{U}{k-1} + \rho\Inv\M{S}) \\ 
\Mn{z}{k} & = & \underset{\M{Z}}{\arg\min}\; \frac{\lambda}{\rho}
\lVert \M{Z} \rVert_1 + \frac{1}{2} \lVert \Mn{B}{k} + \Mn{U}{k-1} -
\M{Z} \rVert_{\text{F}}^2   =  S_{\lambda/\rho}(\Mn{B}{k} +
\Mn{U}{k-1}) \\
\Mn{u}{k} & = & \Mn{u}{k-1} + \Mn{B}{k} - \Mn{Z}{k}.
\end{eqnarray*}
Thus, the penalty and constraint are effectively decoupled resulting
in simple updates: the update for $\M{Z}$ requires the soft-thresholding operator, $S_{\mu}(\x) = \sign{\x}(\lvert \x \rvert
- \mu)_+$, and the update for $\M{B}$ involves the projection onto the
Fantope, denoted by $\mathcal{P}_{\mathcal{F}^d}$, which has a closed
form solution given in \cite{VuChoLei2013}.


The literature abounds with many more examples of using the ADMM
splitting strategy to decouple an otherwise challenging optimization
problem into simpler subproblems.
Boyd et al. \cite{boyd2011distributed} review many such applications.
Other example applications include decoupling trace or nuclear norm
penalties as in robust PCA \cite{YuaYan2012}, latent variable graphical models
\cite{MaXue2013}, and tensor completion
\cite{liu2009tensor}; 
decoupling different types of hierarchical 
constraints \cite{bien2013lasso}, decoupling a series of loss
functions \cite{hu2014local}, decoupling joint graphical models 
\cite{DanWan2014}, and decoupling large  
linear programming problems \cite{AguXinSmi2011}, among many others.


The second thematic challenge that ADMM algorithms have
been used to solve involve fusion or non-separable penalties.  
A good illustrative example of this challenge 
arises in total variation (TV) denoising
\cite{RudOshFat1992}. Consider the simple version of this problem,
specifically finding a smooth estimate of a noisy one-dimensional signal
$y \in \Re^n$: 
\begin{eqnarray*}
\underset{\bbeta}{\text{minimize}}\quad \frac{1}{2} \lVert \V{y} - \bbeta \rVert_2^2 + \lambda \sum_{i=1}^{n-1} \lvert \VE{\beta}{i} - \VE{\beta}{i+1} \rvert,
\end{eqnarray*}
where the tuning parameter $\lambda \geq 0$ trades off the smoothness
of the approximation with the goodness of fit with the data
$\V{y}$. What makes this problem challenging is that the fusion
penalty couples the non-smooth terms so that they are
non-separable. Note that this penalty can be 
written more compactly as $\rVert \mathbf{A} \x \lVert_{1}$ 
where $\M{A}$ is the discrete first order differences operator
matrix. 
More generally, this second class of problems consist of
problems of the form, $L(\bbeta;\mathbf{W})  + \lambda P( \M{A} \bbeta)$. 
In the machine learning context these penalties arise because we often
wish to impose structure, not on a latent variable of interest
directly, but rather on a linear transformation of it. In the TV
denoising example we seek sparsity in differences of adjacent time
points of the latent signal.

Previously, we could break the objective into a sum of simpler
objectives. The issue here is different; specifically the composition
of the regularizer with a linear mapping complicates matters. 
ADMM can again greatly simplify this problem if we let the ADMM copy
variable copy the linearly transformed parameters:
\begin{eqnarray*}
\underset{\bbeta}{\text{minimize}}\quad   L(\bbeta;\mathbf{W}) + \lambda P(\V{z}) \quad \text{subject to}\quad \V{z} - \M{A}\V{\bbeta} = \M{0}.
\end{eqnarray*}
The ADMM subproblems for iteratively solving this problem then have
the following simple form:
\begin{eqnarray*}
\Vn{\bbeta}{k} & = & \underset{\bbeta}{\arg\min} \; \ \ L(\bbeta;\mathbf{W}) + \frac{\rho}{2} \lVert \M{A}\bbeta - \Vn{z}{k-1} + \Vn{u}{k-1} \rVert_2^2 \\
\Vn{z}{k} & = & \underset{\V{z}}{\arg\min}\; \ \ P(\V{z}) + \frac{\rho}{2} \lVert \V{z} - \M{A}\Vn{\bbeta}{k} - \Vn{u}{k-1} \rVert_2^2 \\
\Vn{u}{k} & = & \Vn{u}{k-1} + \Vn{\bbeta}{k} - \Vn{z}{k}.
\end{eqnarray*}
Note that we have eliminated having to minimize any functions
containing the troublesome composition penalty. In the context of the TV
denoising example, the $\bbeta$ update requires solving a linear system
of equations, and the $\V{z}$ update involves a straightforward
soft-threshold.

The ADMM algorithm has been used to decouple fusion or non-separable
types of penalties in many statistical learning problems.  These include
more general instances of total variation 
\cite{wahlberg2012admm,GolOsh2009}, a convex formulation of clustering
\cite{ChiLan2013}, joint graphical model selection 
\cite{MohChuHan2012,MohLonFaz2014}, overlapping group lasso penalties
\cite{yuan2011efficient}, and more generally for structured sparsity
\cite{mairal2011convex}.

Overall, while the ADMM algorithm is gaining more widespread
application in statistics and machine learning, the algorithm is
applied in the traditional sense to solve a composite optimization
problem for one value of the regularization parameter.  In this
chapter, we seek to investigate the ADMM algorithm for a different
purpose, namely to find a sequence of sparse models associated with
regularization paths.

\subsection{Developing an Algorithmic Regularization Path: Sparse
  Regression} 
\label{sec:dev_path}

Our goal is to quickly generate a sequence of candidate
sparse solutions for model selection purposes.  To achieve this, we
will propose a method of approximating the sequence of active sets
given by regularization paths, or the path-like 
sequence solutions of penalized statistical models over the full range of
regularization parameters.  To motivate our approach, we study
the familiar example of sparse linear regression.  Suppose we observe a
covariate matrix $\X \in \Re^{n\times p}$ consisting of $n$ iid
observations of $p$ variables and an associated response
variable $\y \in \Re^n$.  We are interested in fitting the linear model $\y =
\X \bbeta + \epsilon$ where $\epsilon$ is independent noise, but
assume that the linear coefficient vector $\bbeta$ is sparse, $\rVert
\bbeta \lVert_{0} \ll p$ where $\lVert \cdot \rVert_{0}$ is the $\ell_{0}$ norm or
the number of non-zero elements of $\bbeta$. Minimizing a criterion
subject to a constraint 
of the form $\lVert \bbeta \rVert_0 \leq k$ for some $k$, becomes a
combinatorially hard task. 
To estimate a sparse model in reasonable time, many
have proposed to use the tightest convex relaxation, the
$\ell_{1}$-norm penalty, commonly called the LASSO
\cite{tibshiranit1996regression} in the 
statistical literature:  
\begin{align}
\label{eqn:lasso}
\underset{\bbeta}{\text{minimize}}
\quad \frac{1}{2n}\lVert \y - \X \bbeta\rVert_2^2 + \lambda \lVert\bbeta\rVert_1
\end{align}
where $\lambda \geq 0$ is the regularization parameter controlling the
sparsity of $\bbeta$.

The full regularization path of solutions for the LASSO is the set of
regression coefficients 
 $\{ \hat{\bbeta}(\lambda) : \ \forall \ 0 \leq \lambda \leq
\lambda_{\max} \}$ where $\lambda_{\max} = \frac{1}{n} \lVert \X\Tra \y
\rVert_\infty$ is the smallest amount of regularization that yields 
the sparse solution $\hat{\bbeta} = \mathbf{0}$.  The regularization
paths for the LASSO have been well-studied and in particular, are
continuous and piece-wise linear
\cite{osborne2000new,efron2004least,rosset2007piecewise}. 
These paths also outline a 
sequence of active sets or sparse models that smoothly increase in
sparsity levels as $\lambda$ decreases from the fully sparse solution at
$\lambda = \lambda_{\max}$ to the fully dense solution at $\lambda =
0$.  Hence for model selection, one can limit exploration of the
combinatorial space of sparse models to that of the sequence of active
sets outlined by the LASSO regularization paths.

Computing the full regularization paths, however, can be a
computational challenge.  Several path following algorithms for the
LASSO \cite{osborne2000new,rosset2007piecewise} and closely related
algorithms such as such as Least 
Angle Regression (LAR) \cite{efron2004least} and Orthogonal 
Matching Pursuit (OMP) \cite{donoho2012sparse} have been
proposed; their computational complexity, however, is $\mathcal{O}(p^{3})$ which
is prohibitive for large-scale problems.  Consequently, many have
suggested to closely approximate these paths by solving a series of
optimization problems over a grid of regularization parameter values. 
Specifically, this is typically done for a sequence of 100 log-spaced
values from $\lambda_{\max}$ to $\lambda_{1} = 0$.
Statisticians often employ homotopy, or warm-starts, to speed
computation along the regularization path \cite{friedman2007pathwise}.
Warm-starts use 
the solution from the previous value of $\lambda_{j}$,
$\hat{\bbeta}(\lambda_{j})$, as the initialization for the optimization
algorithm to solve the problem at $\lambda_{j+1}$.  As the
coefficients, $\bbeta$, change continuously in $\lambda$, warm-starts
can dramatically reduce the number of iterations needed for
convergence as $\hat{\bbeta}(\lambda_{j})$ is expected to be close to
$\hat{\bbeta}(\lambda_{j+1})$ for small changes from $\lambda_{j}$ to
$\lambda_{j+1}$.  Many consider shooting methods, or coordinate
descent procedures \cite{friedman2007pathwise,wu2008coordinate}, that
use warm-starts and iterate over the 
active set for 100 log-spaced values of $\lambda$
\cite{friedman2010regularization} to be the fastest 
approximate solvers of the LASSO regularization path.


We seek to further speed the computation of the sequence of active
sets given by the regularization path by 
using a single path approximating algorithm instead of solving
separate optimization problems over a grid of regularization parameter
values.  Our approach is motivated by two 
separate observations: (i) the evolution of the sparsity level of the
iterates of the 
ADMM  
algorithm used to fit \eqref{eqn:lasso} for one value of $\lambda$, and (ii)
the behavior of a new version of the ADMM algorithm that
incorporates warm-starts to expedite computation of regularization
paths.  We study each of these motivations separately, beginning
with the first.

Consider using ADMM to solve the LASSO
problem.  First, we 
split the differentiable loss function from the non-differentiable
penalty term by introducing a copy $\z$ of the variable $\bbeta$ in the penalty
function, and adding an equality constraint forcing them to be equal.
The LASSO problem \eqref{eqn:lasso} can 
then be re-written as: 
\[\underset{\bbeta,\z}{\text{minimize}}
\quad \frac{1}{2n}\lVert  \y - \X \bbeta\rVert_2^2 + \lambda \lVert\z\rVert_1
\quad \text{subject to} \quad \bbeta = \z,
\]
with its associated augmented Lagrangian:
\[
\mathcal{L}(\bbeta,\z,\uu) =  \frac{1}{2n}\lVert \y - \X \bbeta\rVert_2^2 + \lambda \lVert\z\rVert_1 + \frac{\rho}{2}\lVert \bbeta - \z + \uu  \rVert_2^2.
\]
Here, $\uu$ is the scaled dual variable of the same dimension as
$\bbeta$ and $\rho$ is the algorithm tuning parameter.  The ADMM
algorithm then follows three steps 
(subproblems) to solve the LASSO: 
 \[
 \begin{aligned}
&\bbeta\text{-subproblem:  } \bbeta^k = \underset{\bbeta}{\arg\min}\;  \frac{1}{2n}\lVert \y - \X \bbeta\rVert_2^2 + \frac{\rho}{2}\lVert \bbeta - \z^{k-1} + \uu^{k-1}  \rVert_2^2 \\
&\z\text{-subproblem:  } \z^k = \underset{\z}{\arg\min}\; \lambda \lVert\z\rVert_1+ \frac{\rho}{2} \lVert  \z-\bbeta^k- \uu^{k-1}  \rVert_2^2\\
&\text{Dual update:  } \uu^k = \uu^{k-1} + \bbeta^k-\z^k.
\end{aligned}
\]
The benefit of solving this reformulation is simpler iterative updates.
These three steps are iterated until convergence, typically measured
by the primal and dual residuals \cite{boyd2011distributed}.  The
$\bbeta$-subproblem 
solves a linear regression with an 
additional quadratic ridge penalty.  Solving the $\z$-subproblem
introduces sparsity.  Notice that this is the
proximal operator of the $\ell_{1}$-norm applied to $\bbeta^k -
\uu^k$ which is solved analytically via soft-thresholding.
  Finally, the dual update
ensures that $\bbeta$ is squeezed towards $\z$ and primal feasibility as the
algorithm progresses.

\begin{figure}[ht] 
   \centering
   \includegraphics[width=\textwidth,clip=true,trim=1.2in 0in 1.1in 0in]{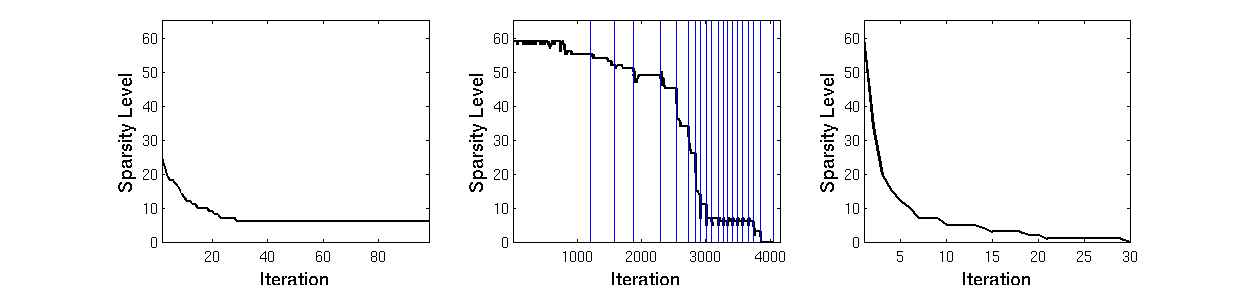} 
   \caption{Sparsity levels outlining a sequence of active
     sets for a simulated sparse linear regression example. (Left)
     Sparsity levels of the $\z$-subproblem 
   iterates of the ADMM algorithm, $\lVert \z^k \rVert_{0}$, fit for one
   fixed value of 
   $\lambda$.  (Middle) Sparsity levels of the $\z$-subproblem over
   the iterates of our path 
   approximating ADMM
   Algorithm with Warm Starts for a small range of 
   $\boldsymbol{\lambda}$.  Vertical lines denote the start of the
   algorithm for an increased value of $\lambda$.  (Right) Sparsity
   levels of the $\z$-subproblem over the iterates of our novel ADMM
   Algorithmic Regularization Path.}
   \label{fig:admm_sparsity}
\end{figure}

Consider the sparsity of the $\z$ iterates, $\lVert \z^k \rVert_0$, for the
LASSO problem.  Notice that as the algorithm proceeds, $\z^k$ becomes
increasingly sparse; this is illustrated for a small simulated 
example in the left panel of Figure~\ref{fig:admm_sparsity}.  Let us
study why this occurs and its implications.  Regardless of
$\lambda$, the ADMM algorithm begins with a fully dense $\bbeta^1$ as
this is the solution to a ridge problem with parameter $\rho$.
Soft-thresholding in the $\z$-subproblem then sets coefficients of
small magnitude to zero.  The first dual update, $\uu^1$, has
magnitude at most $| \lambda|$, meaning that the second $\bbeta^2$
update is essentially shrunken towards $\bbeta^1$.  Smaller
coefficients decrease further in magnitude and soft-thresholding in
the $\z$-subproblem sets even more coefficients to zero.  The
algorithm thus proceeds until the sparsity of the $\z^k$
stabilizes to that of the solution, $\hat{\bbeta}(\lambda)$.  Hence,
the support of the $\z^k$ has approximated the active set of
the solution long before the iterates of the $\bbeta$-subproblem; the
latter typically does not reach the sparsity of the solution until
convergence when primal feasibility is achieved.  While
Figure~\ref{fig:admm_sparsity} only illustrates that $\z^{k}$ quickly
converges to the correct sparsity level, we have observed empirically
in all our examples that the active set outlined by $\lVert \z^k
\rVert_0$ also quickly identifies the true non-zero elements of the
solution, $\hat{\bbeta}(\lambda)$.

Interestingly then, the $\z^k$ quickly explore a sequence of sparse models going
from dense to sparse, similar in nature to the sequence of sparse
models outlined by the regularization path.  While from
Figure~\ref{fig:admm_sparsity} 
we can see that this sequence of sparse models is not desirable as
it does not smoothly transition in sparsity and does not fully explore
the sparse model space, we nonetheless learn  two important
items from this: (i) We are motivated to

consider using the algorithm iterates of
the $\z$-subproblem, as a possible means of quickly exploring the sparse model
space; and (ii) we are motivated to consider a sequence of 
solutions going from dense to sparse as this naturally aligns with the
sparsity levels observed in the ADMM algorithm iterates.
Given these, we ask: Is it possible to use or modify the iterates of the
ADMM algorithm to achieve a path-like smooth transition in sparsity
levels similar in nature to the sparsity levels and active sets 
corresponding to regularization paths?

One possible solution would be to employ warm-starts in the ADMM
algorithm along a grid of regularization parameters similar to other
popular algorithms for approximating 
regularization paths.  Recall that warm-starts use the solution
obtained at the previous value of $\lambda$ as initialization for the
next value of $\lambda$.  We first introduce this new extension of the
ADMM algorithm for approximating regularization paths in 
Algorithm~\ref{alg_admm_warm} and then return to our motivation of
studying the sequence of active sets outlined by this algorithm.
\begin{algorithm}
\caption{ADMM with Warm Starts: Sparse Regression}
\label{alg_admm_warm}
\begin{enumerate}
\item[1.] Initialize  $\bbeta^0 = \0$, $\uu^0 = \0$, and $M$ log-spaced
  values, $\boldsymbol{\lambda} = \{ \lambda_{1} < \lambda_2 < \ldots <
  \lambda_{M} \}$, for $\lambda_{1} = 0$ and $\lambda_{M} =
  \lambda_{\max}$.  
\item[2.] Precompute matrix inverse $\Hb = (\X\Tra\X/n + 
  \Ib)\Inv$ and $\Hb \X\Tra\y$.
\item[3.] 
\begin{algorithmic}[0]
  \For{$j = 1 \ldots M$}
  \While{ $\lVert \mathbf{r}^k \rVert \wedge \lVert \mathbf{s}^k \rVert > \epsilon^{\text{tol}}$}
  \State 
  $\z_j^k = S_{\lambda_{j}}(\bbeta_j^{k-1}+\uu_j^{k-1})$
  \State 
  $\bbeta_j^k = \Hb \X\Tra\y+ \Hb(\z_j^{k} - \uu_j^{k-1})$
  \State 
  $\uu_j^k = \uu^{k-1} + \bbeta_j^k-\z_j^k$
  \State 
  $\mathbf{r}^k = \bbeta_j^k - \z_j^k$ and $\mathbf{s}^k = \z_j^k -
  \z_j^{k-1}$
  \State
  $k = k + 1$
  \EndWhile
  \EndFor
\end{algorithmic}
\item[4.]  Output $\{\bbeta_j :j = 1,\cdots, M \}$ as the regularization
  path. 
\end{enumerate}
\end{algorithm}

Our ADMM algorithm with warm starts is an alternative algorithm for
fitting regularization paths. It begins with $\lambda$ small
corresponding to a dense model, fits the ADMM algorithm to obtain the
solution, and then uses the previous solution, $\bbeta(\lambda_{j-1})$, and
dual variable, $\uu(\lambda_{j-1})$, to initialize the ADMM algorithm for
$\lambda_{j}$. 

Before considering the sequence of active sets
outlined by this algorithm,
we pause to discuss some noteworthy features.
First, notice that the ADMM tuning parameter, $\rho$, does not appear
in this algorithm.  We have omitted this as a parameter by fixing
$\rho = 1$ throughout the algorithm.  Fixing $\rho$ stands in contrast
with the burgeoning literature on how to dynamically update $\rho$ for 
ADMM algorithms \cite{boyd2011distributed}. For example, adaptive
procedures that change $\rho$ to speed up convergence are introduced
in \cite{he2000alternating}.  Others have proposed accelerated
versions of the ADMM algorithm that achieve a similar phenomenon
\cite{goldstein2012fast}.  Changing the algorithm tuning parameter,
however, is not conducive to achieving a path-like algorithm using
warm-starts.  Consider the $\z$-subproblem which is solving by
soft-thresholding at the level $\lambda_{j} / \rho$.  Thus, if $\rho$
is changed in the algorithm, the sparsity  levels of $\z$ dramatically
change, eliminating the advantages of using warm-starts to
achieve smooth transitions in sparsity levels.  
Second, notice that we have switched the order of the sub-problems by
beginning with the $\z$-subproblem.  While technically the order of
the subproblems does not matter \cite{yan2014self}, we begin with the $\z$-subproblem as
this is where the sparsity is achieved through soft-thresholding at
the value, $\lambda$; hence, the solution for $\z$ is what changes when
$\lambda$ is increased.

Next, notice that our regularization paths go from dense to sparse, or
$\lambda$ small to large, which is the opposite of other path-wise
algorithms and algorithms that approximate regularization paths over a
grid of $\lambda$ values \cite{friedman2010regularization}.  Recall
that our objective is to 
obtain a smooth path-like transition in sparsity levels corresponding
to a sequence of active sets that fully
explores the space of sparse models.  Our warm-start procedure
naturally aligns with the sparsity levels of the iterates of the ADMM
algorithm which go from dense to sparse, thus ensuring a smooth
transition in the sparsity level of $\z$ as $\lambda$ is increased.
Our warm-start procedure could certainly be employed going in the
reverse direction from sparse to dense, but we have observed that this
introduces discontinuities in the $\z^k$ iterates and
consequently 
their active sets as well, thus 
requiring more iterations for convergence.   This behavior occurs
as the solution of the $\bbeta$-subproblem is always more dense than that of
the $\z$-subproblem, even when employing warm-starts.

Now, let us return to our motivation and consider the sparsity levels
and corresponding sequence of active sets achieved by the iterates of
our new path-approximating ADMM Algorithm.  The sparsity of the
iterates of the $\z$-subproblems, $\lVert \z^k \rVert_{0}$, are
plotted for 30 
log-spaced values of $\lambda$ for the 
same simulated example in the middle panel of
Figure~\ref{fig:admm_sparsity}.   The iterates over all values of
$\lambda$ are plotted on the x-axis with 
vertical lines denoting the increase to the next $\lambda$ value.
Carefully consider the sparsity levels of the $\z$ iterates for each
fixed value of $\lambda$ in our ADMM algorithm with warm starts.
Notice that the sparsity levels of $\z$ typically stabilize to that of
the solution within the first few iterations after
$\lambda$ is increased.  The remaining iterations and a large
proportion of the computational time are spent on
squeezing $\bbeta$ towards $\z$ to satisfy primal feasibility.  This
means that the $\z$-subproblem has achieved the sparsity
associated with the active set of $\hat{\bbeta}(\lambda)$ within a few
iterations of increasing $\lambda$.  One could surmise that if the
increase in $\lambda$ were small enough, then the $\z$-subproblem
could correctly approximate the active set corresponding to $\lambda$
within one iteration when using this warm-start procedure.  The right
panel of Figure~\ref{fig:admm_sparsity} illustrates the sparsity
levels achieved by the $\z$-subproblem for this sequence of one-step
approximations to our ADMM algorithm with warm-starts.  Notice that
this procedure achieves a smooth transition in sparsity levels
corresponding to a sequence of active sets that fully explore the
range of possible sparse models, but requires only a fraction of the
total number of iterations and compute time.  This, then 
is the motivation for our new ADMM Algorithmic Regularization Paths
introduced in the next section.

\section{The Algorithmic Regularization Path}
\label{sec:admm_path_general}

Our objective is to use the ADMM splitting method as the foundation
upon which to develop a new approximation to the sequence of sparse
solutions outlined by regularization paths.  In doing so, we are not
interested in estimating parameter values by solving a statistical
learning optimization problem with high precision.  Instead, we are
interested in quickly exploring the space of sparse
model at a fine resolution for model selection purposes by
approximating the sequence of active sets given by the regularization
path.

Again, consider the general sparse statistical machine learning problem of the
following form:
\[
\underset{\bbeta}{\text{minimize}}\quad L(\bbeta;\mathbf{W}) + \lambda P(\bbeta),
\]
where $\mathbf{W}$ denotes the ``data'' (for the sparse linear
regression example, $\mathbf{W} = \{ \X, \y \}$), the loss
function, $L( \bbeta; \mathbf{W})$ is a differentiable, convex function of
$\bbeta$, and the regularization term, $P: \Re^{p} \rightarrow
\Re^{+}$ is a convex and non-differentiable penalty function.  As
before, $\lambda \geq 0$ is the regularization parameter controlling
the trade-off between the penalty and loss function. Following the
setup of the ADMM algorithm, consider splitting the smooth loss from
the nonsmooth penalty through the copy variable, $\z$:
\begin{equation}
\label{eq:general}
\underset{\bbeta,\z}{\text{minimize}}\quad L(\bbeta; \mathbf{W}) + \lambda P(\z)\quad \text{subject to } \bbeta = \z,
\end{equation}
With scaled dual variable $\uu$, the augmented Lagrangian of  general problem \eqref{eq:general} is 
\[
\mathcal{L}(\bbeta,\z,\uu) = L(\bbeta; \mathbf{W}) +\lambda P(\z) + \frac{\rho}{2}\lVert \bbeta - \z + \uu  \rVert_2^2.
\]


Now following from the motivations discussed in the previous section,
there are three key ingredients that we employ in our Algorithmic
Regularization Paths: 
(i) warm-starts to go from a dense to a sparse solution, (ii) the
sparsity patterns of the $\z$-subproblem iterates, and (iii) one-step
approximations at each regularization level.  We put these ingredients
together in Algorithm~\ref{alg_gen} to give our Algorithmic
Regularization Paths:
\begin{algorithm}
\caption{Algorithmic Regularization Path for Sparse Statistical Learning}
\label{alg_gen}
\begin{enumerate}
\item[1.] Initialize  $\z^0 = \0$, $\uu^0 = \0$, $\gamma^0 =
  \epsilon$, $k = 1$, and set $t>0$. 
\item[2.] While $\lVert\z^k\rVert\neq 0 $
\newline \hspace*{.2in} $\gamma^k = \gamma^{k-1}+t$ (or $\gamma^k =
\gamma^{k-1}t$)
\newline \hspace*{.2in} $\bbeta^k = \underset{\bbeta}{\text{minimize}}\quad L(\bbeta;\mathbf{W}) + \frac{1}{2}\lVert \bbeta - \z^{k-1} + \uu^{k-1}  \rVert_2^2$
\newline \hspace*{.2in} $\z^k = \underset{\z}{\text{minimize}}\quad
\gamma^k P(\z)+ \frac{1}{2}\lVert  \z-\bbeta^k- \uu^{k-1}
\rVert_2^2$ \hspace{.1in}  (Record $\z^k$ at each iteration.)
\newline \hspace*{.2in} $\uu^k = \uu^{k-1} + \bbeta^k-\z^k$
\newline \hspace*{.2in} $k = k+1$
\newline end
\item[3.]  Output $\{\z^k:k = 1,\cdots, K\}$ as algorithmic
  regularization path . 
\end{enumerate}
\end{algorithm}

Our Algorithmic Regularization Path, Algorithm Path for short,
outlines a sequence of sparse models going from fully dense to fully
sparse.  This can be used as an approximation to the sequence of
active sets given by regularization paths for the purpose of model
selection.  Consider that the algorithm begins will the fully dense ridge
solution.  It then gradually increases the amount of regularization,
$\gamma$, performing one full iterate of the ADMM algorithm
($\bbeta$-subproblem, $\z$-subproblem, and dual update) for each new
level of regularization.  The regularization level is increased until
the $\z$-subproblem becomes fully sparse.

One may ask why we would expect our Algorithm Path to yield a sequence
of active sets that well approximate those of the regularization path?
While a mathematical proof of this is beyond the scope of this
chapter, we outline the intuition stemming from our three key
ingredients.  (Note that we also demonstrate this through specific
examples in the next section).  
\begin{itemize}
\item[(i)] Warm-starts from dense to sparse.  Beginning with a dense
  solution and gradually increasing the amount of regularization
  ensures a smooth decrease in the sparsity levels corresponding to a
  smooth pruning of the active set as this naturally aligns with
  sparsity levels of the ADMM algorithm iterates.
\item[(ii)] $\z$-subproblem iterates.  The iterates of the
  $\z$-subproblem encode the sparsity of the active set,
  $\hat{\bbeta}(\lambda)$, quickly as compared to the
  $\bbeta$-subproblem which achieves sparsity only in the limit upon
  algorithm convergence. 
\item[(iii)] One-step approximations.  For a small increase in
  regularization when using warm-starts, the iterates of the
  $\z$-subproblem often achieve the sparsity level of the active set
  within one-step. 
\end{itemize}
Notice that if we iterated the three subproblems of our Algorithm Path
fully until convergence, then our algorithm would be equivalent to our
ADMM Algorithm with warm starts; thus, the one-step approximation is
the major difference between Algorithms \ref{alg_admm_warm} and
\ref{alg_gen}.  Because of this one-step approximation, we
are not fully solving \eqref{eq:the_problem} and thus the parameter
values, $\bbeta$, will never stabilize to that of the regularization
path.  Instead, our Algorithm Path quickly approximates the sequence
of active sets corresponding to the regularization path, as encoded in
the $\z$-subproblem iterates.


The astute reader will notice that we have denoted the regularization
parameters in Algorithm~\ref{alg_gen} as $\gamma$ instead of
$\lambda$ as in \eqref{eq:the_problem}.  This was intentional since
due to the one-step approximation, we are not solving
\eqref{eq:the_problem} and thus the level of regularization achieved,
$\gamma$, does not correspond to $\lambda$ from
\eqref{eq:the_problem}.  Also notice that we have introduced a step
size, $t$, that increases the regularization level,
$\gamma$, at each iteration.  The sequence of $\gamma$'s can either be
linearly spaced, as with additive $t$, or geometrically spaced, as
with multiplicative $t$.  Again, if $t$ is very small, then we expect
the sparsity patterns of our Algorithm Paths to well approximate the
active sets of regularization paths.

We will explore the behavior and benefits of our Algorithm Paths through
demonstrations on popular sparse statistical learning problems
in the next section.  Before presenting specific examples, however, we
pause to outline three important advantages that are general to sparse
statistical learning problems of the form \eqref{eq:the_problem}.
\begin{enumerate}
\item Easy to implement.  Our Algorithm Path is much simpler than
  other algorithms to approximate regularization paths.  The hardest
  parts, the $\bbeta$ and $\z$ subproblems, often have analytical
  solutions for many popular statistical learning methods.  Then, with
  only one loop, our
  algorithm can often be implemented in a few lines of code.
  This is in contrast to other algorithm paths which require multiple
  loops and much
  overhead to track algorithm convergence or the coordinates of active
  sets \cite{friedman2010regularization}.   
\item Finer resolution exploration of the model space.  Our Algorithm
  Path has the potential to explore the space of sparse models at a
  much finer resolution than other fast methods for approximating
  regularization paths over a grid of $\lambda$ values.  Consider that
  as the later are computed over $M$, typically $M = 100$, $\lambda$
  values, these can yield an upper bound of $M$ distinct active sets;
  usually these yield much less than $M$ distinct models.  In
  contrast, our Algorithm Path yields an upper bound of $K$ distinct
  models where $K$ is the number of iterations needed, depending on
  the step-size $t$, to fully explore the sequence of sparse models.
  As $K$ will often be much greater than $M$, our Algorithm Path will
  often explore a sequence of many more active sets and at a finer
  resolution than comparable methods. 
\item Computationally fast.  Our Algorithm Path has the potential to
  yield a sequence of sparse solutions much faster than other methods
  for computing regularization paths.  Consider that our algorithm
  takes at most $K$ iterations.  In contrast, regularization paths of a
  grid of $M$ $\lambda$ values require $M$ times the number of
  iterations needed to fully estimate $\hat{\bbeta}(\lambda_j)$; often
  this will be much larger than $K$.  In each iteration of our
  algorithm, the $\bbeta$ and $\z$ subproblems require the most
  computational time.  The $\bbeta$
  subproblem consists of the loss function with a quadratic penalty
  which can be solved via an analytical form for many loss functions.
  The $\z$ subproblem has the form of the proximal operator of $P$
  \cite{parikh2013proximal}: $\mathrm{prox}_{\lambda P}(\x) = \arg\min_{\uu}
  \lVert \x - \uu\rVert_{2}^{2} + \lambda
  P(\uu)$.   For many popular convex penalty-types 
  such as the $\ell_{1}$-norm, group lasso, and nuclear norm, the
  proximal operator has an analytical solution.  Thus, for a large
  number of statistical machine learning problems, the iterations of
  our Algorithm Path are inexpensive to compute.
\end{enumerate}

Overall, our Algorithmic Regularization Paths give a novel method for
finding a sequence of sparse solutions by approximating the active
sets of regularization paths.  Our methods can be used in place of
regularization paths for model selection purposes with many sparse
statistical learning problems.  In this chapter, instead of studying
the mathematical and statistical properties of our new Algorithm
Paths, which we leave for future work, we study our method through
applications to several statistical learning problems in the next
section.

\section{Examples}
\label{sec:examples}

To demonstrate the versatility and advantages of our ADMM
Algorithmic Regularization Paths, we present several example
applications to sparse statistical learning methods: sparse linear
regression, reduced-rank multi-task learning and convex clustering.  

\subsection{Sparse Linear Regression}
\label{sec:admm_path_lasso}

As our first example, we revisit the motivating example of sparse
linear regression discussed in
Section~\ref{sec:dev_path}.  We reproduce the problem here for convenience:
\begin{align*}
\underset{\bbeta}{\text{minimize}}
\quad \frac{1}{2n}\lVert \y - \X \bbeta\rVert_2^2 + \lambda \lVert\bbeta\rVert_1
\end{align*}
And, our Algorithmic Regularization Path for this example is presented
in Algorithm~\ref{alg_lasso}:
\begin{algorithm}
\caption{Algorithmic Regularization Path for Sparse Regression}
\label{alg_lasso}
\begin{enumerate}
\item[1.] Initialize  $\z^0 = \0$, $\uu^0 = \0$, $\gamma^0 =
  \epsilon$, $k=1$, and set $t>0$.
\item[2.] Precompute matrix inverse $\Hb = (\X\Tra\X/n + 
  \Ib)\Inv$ and $ \Hb \X\Tra\y$.
\item[3.] While $\lVert\z^k\rVert \neq 0$
\newline \hspace*{.2in} $\gamma^k = \gamma^{k-1} + t$
\newline \hspace*{.2in} $\bbeta^k = \Hb \X\Tra\y+ \Hb(\z^{k-1} - \uu^{k-1})$
\newline \hspace*{.2in} $\z^k =
S_{\gamma^k}(\bbeta^k+\uu^{k-1})$ \hspace{.1in} (Record $\z^k$ at
each iteration.) 
\newline \hspace*{.2in} $\uu^k = \uu^{k-1} + \bbeta^k-\z^k$
\newline \hspace*{.2in} $k = k+1$
\newline end
\item[4.]  Output $\{\z^k:k = 1,\cdots, K\}$ as the algorithmic
  regularization path . 
\end{enumerate}
\end{algorithm}

Let us first discuss computational aspects of our Algorithm Path for
sparse linear regression.  Notice that the $\bbeta$-subproblem
consists of solving a ridge-like regression problem.  Much of the
computations involved, however, can be pre-computed, specifically the
matrix inversion, $(\X\Tra\X/n+\Ib)^{-1}$, and matrix-vector multiplication, 
$\X\Tra \y$.  In cases where $p \gg n$, inverting a $p \times p$ matrix
is is highly computationally intensive, requiring
$\mathcal{O}(p^3)$ operations.  We can reduce the computational cost to
$\mathcal{O}(n^3)$, however, by invoking the Woodbury Matrix Identity
\cite{hager1989updating}:  $\left( \X\Tra\X/n+\Ib_p \right )\Inv  = 
\Ib_{p} - \X\Tra (n\Ib_n + \X \X\Tra)\Inv\X$ and caching the Cholesky
decomposition of the smaller 
$n$-by-$n$ matrix $n\Ib_n + \X\X\Tra$.
Thus, the iterative updates for $\bbeta^k$ are reduced to
$\mathcal{O}(n^2)$, the cost of solving 
two $n$-by-$n$ triangular linear systems of equations. The $\z$-subproblem is
solved via soft-thresholding
which requires only $\mathcal{O}(p)$ operations.

\begin{figure}[ht] 
   \centering
\includegraphics[width=\textwidth]{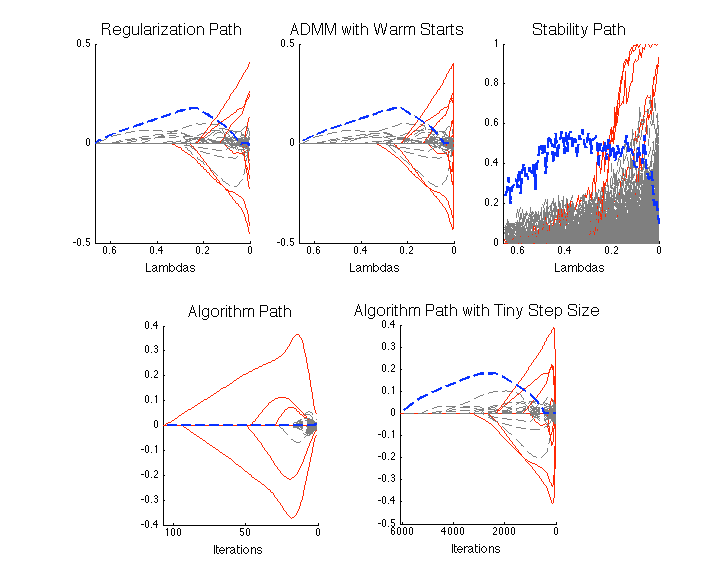}
   \caption{Comparisons of Algorithmic Regularization Paths (bottom
     panel) to
     regularization paths (top left and middle) and stability paths
     (top right) for the sparse linear regression example.
     The {\color{gray}$--$}
     lines denote false variables, {\color{red}\textemdash} lines
     denote true non-zero variables, and  {\color{blue}$--$} lines denote some 
     highlighted false positives.
     Regularization paths were computed via the popular shooting
     method~\cite{friedman2010regularization} (top left) and our ADMM
     algorithm with warm-starts 
     (top middle).  Our Algorithmic Regularization Path with a tiny
     step size (bottom right) closely approximates the sparsity patterns of
     the regularization
     paths, while our method with a larger step size (bottom left)
     dramatically differs from the regularization paths.  Notice that
     sparse regression in this example does a poor job of variable
     selection, selecting many false positives before any true
     features enter the model.  Even the stability paths (top right)
     select many false positives.  Our Algorithmic Regularization Path
     with a larger step-size, however, selects many of the true
     variables with much fewer false positives.}
   \label{fig:path_comp2}
\end{figure}

We study our Algorithmic
Regularization Path for sparse linear regression through a real data
example.  We  
use the publicly available 14-cancer microarray data from
\cite{hastie2009elements} 
to form our covariate matrix.  This consists of gene expression
measurements for
$n = 198$ subjects and 16063 genes; we randomly
sample $p = 2000$ genes to use as our data matrix $\X$.  We simulate
sparse true 
signal $\bbeta^*$ with $s = 16$ non-zero features of absolute
magnitude 5-10, and with the
signs of the non-zero signals assigned randomly; the 16
non-zero variables were randomly chosen from the 2000 genes.  The
response variable $\y$ is generated as $\y = \X \bbeta^* +
\mathbf{\epsilon}$, where $\epsilon
\stackrel{i.i.d.}{\sim}N(0,1)$. A visualization of regularization
paths, stability paths, and our Algorithmic Regularization Paths is
given in Figure~\ref{fig:path_comp2} for this example.

First, we verify empirically that that our ADMM algorithm with warm
starts is equivalent to the regularization path (top left and top
middle).  Additionally, notice
that, as expected, our Algorithm Path with a tiny step size (bottom
right) also well
approximates the sequence of active sets given by the
regularization paths.  With a larger step-size, 
however, our algorithm path (bottom left) yields a sequence of sparse
models that differ markedly from the sparsity
patterns of the regularization paths.  This occurs as the change in
regularization levels of each step are large enough so that
the sparsity levels of the $\z$-subproblem after the one-step
approximation are not equivalent to that of the solution to
\eqref{eqn:lasso}.

Despite this, Figure~\ref{fig:path_comp2}
suggests that our Algorithm Paths with larger step sizes
may have some advantages in terms of variable selection.  Notice that
regularization paths select many false positives (blue and gray dashed
lines) before the true positives (red lines).  This is expected as we used a
real microarray data set for $\X$ consisting of strongly correlated
variables that directly violate the irrepresentable conditions under
which variable selection for sparse regression is achievable
\cite{buhlmann2011statistics}.   Our method, however,
selects several true variables before the first false positive enters
the model.  To understand this further, we compare our approach to the
Stability Paths used for stability
selection~\cite{meinshausen2010stability}, a re-sampling 
scheme with some of the strongest
theoretical guarantees for variable selection.  The stability paths,
however, also select several false positives.  This as well as other
empirical results that are omitted for space reasons suggest that
our Algorithm Path with moderate or larger step sizes may perform
better than convex optimization 
techniques in terms of variable selection.  While a theoretical
investigation of this is beyond the scope of this book chapter, the
intuition for this is readily apparent.  Our Algorithm Path starts
from a dense solution and uses a ridge-like penalty.  Thus,
coefficients of highly correlated variables are likely to be grouped
and have similar magnitude coefficient values.  When soft-thresholding
is performed in the $\z$-subproblem, variables which are strongly
correlated are likely to remain together in the model for at least the
first several algorithm iterations.  By keeping correlated variables
together in the model longer and otherwise eliminating irrelevant
variables, this gives our algorithm a better chance of selecting the
truly non-zero variables out of a correlated set.  Hence, the fact
that we start with a dense solution seems to help us; this is in
contrast to the LASSO, LAR and OMP paths which are initialized with an
empty active set 
and greedily add variables most correlated with the
response \cite{osborne2000new,efron2004least}.  We plan on
investigating our methods in terms of 
variable selection in future work.

\begin{table}
\center
\caption{Timing comparison (averaged over 50 replications) of our ADMM
  Algorithmic Regularization Paths, Regularization Paths obtained from
  the shooting method (coordinate descent) \& Stability Paths for
  different numbers of variables in the true model. }
\label{table:timing}
\hspace*{-.2in}\begin{tabular}{ l | c|c|c}
\hline\hline
 Time (seconds) &Algorithmic Regularization Path&Regularization Path &Stability Path \\
\hline 
           s = 20, p = 4000&    0.0481&     0.1322&   36.6813\\
           s = 20, p = 6000&   0.0469&0.1621&43.9320 \\
\hline\hline
 \end{tabular}
\end{table}

Finally, we compare our Algorithm Paths to state-of-the-art methods
for computing the sparse regression regularization paths in terms of
computational time in Table~\ref{table:timing}.  The regularization paths were
computed using the {\tt glmnet} {\tt R} package
\cite{friedman2010regularization} which is based on 
shooting (coordinate descent)
routines~\cite{friedman2010regularization}.  This approach and
software is 
widely regarded as one of the fastest solvers for sparse regression.
Notice that our Algorithm Paths, coded entirely in Matlab, run in
about a fifth of the time as this state-of-the-art competitor.   Also,
our computational time is far superior to the re-sampling schemes
required to compute the stability paths.

Overall, our Algorithmic Regularization Path for sparse linear
regression reveals major computational advantages for finding a
sequence of sparse models that approximate the active sets of
regularization paths.  Additionally, empirical
evidence suggests that our methods may also enjoy some important
statistical advantages in terms of variable selection that we will
explore in future work.

\subsection{Reduced-Rank Multi-Task Learning}
\label{sec:admm_path_rrr}

Our ADMM Algorithmic Regularization Path applies generally to many
convex penalty types beyond the $\ell_{1}$-norm.  Here, we demonstrate
our method in conjunction with a reduced-rank multi-task learning
problem also called multi-response regression.  This problem has been
studied by \cite{negahban2011estimation} among many others.  

Suppose we observe $n$ 
iid samples measured on $p$ covariates and for $q$ outcomes, yielding
a covariate matrix, $\X \in \Re^{n \times p}$, and a response matrix
$\Y \in \Re^{n \times q}$.  Then, our goal is to fit the following
statistical model: $\Y = \X \B + \mathbf{\epsilon}$, where $\B$ is the
$p \times q$ coefficient matrix which we seek to learn, and
$\mathbf{\epsilon}$ is independent noise.  As often the number of
covariates is large relative to the sample size, $pq \gg n$, many have
suggested to regularize the coefficient matrix $\B$ by assuming it has
a low-rank structure, $\mathrm{rank}(\B) < p \wedge q$.  Thus, our
model space of sparse solutions is given by the space of all possible
reduced-rank solutions.  Exploring this space is an NP hard
computational problem; thus,
many have employed the nuclear norm penalty, $\lVert \B \rVert_{*} =
\sum_{j=1}^{p \wedge q} \sigma_{j}( \B)$, which is the sum (or
$\ell_{1}$-norm) of the singular values of $\B$, $\sigma(\B)$, and the
tightest convex relaxation of the rank constraint.  Thus,
we arrive at 
the following optimization problem:
\begin{align}
\label{eq:rrr}
\underset{\B}{\text{minimize}}
\quad \frac{1}{2}\lVert \Y - \X \B\rVert_{\text{F}}^2 + \lambda \lVert\B\rVert_*
\end{align}
Here, $\lVert \cdot \rVert_{\text{F}}$ is the Frobenious norm, $\lambda \geq 0$ is
the regularization parameter controlling the rank of the solution and
$\lVert \cdot \rVert_{*}$ is the nuclear norm penalty.

For model selection then, one seeks to explore the sequence of
low-rank solutions obtained as $\lambda$ varies.  To develop our
Algorithm Path for approximating this sequence of low-rank solutions,
let us consider the ADMM sub-problems for solving \eqref{eq:rrr}.
The augmented Lagrangian, sub-problems, dual updates are analogous 
to that of the sparse linear regression example, and hence we omit
these here. 
Examining the $\Z$-subproblem, however, recall that this is the
proximal operator for the nuclear norm penalty: 
$\Z^k  =  \underset{\Z}{\arg\min}
\; \frac{1}{2}\lVert \Z - ( \B^{k} + \U^{k})\rVert_{\text{F}}^2 + \gamma \lVert\Z\rVert_*$,
which can be solved by soft-thresholding the singular values: Suppose
that $\mathbf{A} = \U \boldsymbol{\Sigma} \mathbf{V}\Tra $ is the SVD of $\mathbf{A}$.  Then
singular-value thresholding is defined as $\text{SVT}_{\gamma}(\mathbf{A}) = \U
[ \mathrm{diag}( (\boldsymbol{\sigma} - \gamma)_{+} )] \mathbf{V}\Tra$ and the
solution for the $\Z$ sub-problem is  $\Z^k = \text{SVT}_{\gamma}(\B^{k} + \U^{k})$.

\begin{algorithm}
\caption{Algorithmic Regularization Path for Reduced-Rank Regression}
\label{alg_rrr}
\begin{enumerate}
\item[1.] Initialize: $\Z^0 = \mathbf{0}$, $\U^0 = \mathbf{0}$,
  $\gamma^0 = \epsilon$, and step size $t > 0$.   
\item[2.] Precompute: $\Hb = (\X\Tra\X/n +  \Ib)\Inv$ and $\Hb
  \X\Tra \Y$.
\item[3.] While $\lVert \Z^k \rVert \neq 0$
\newline \hspace*{.2in} $\gamma^k = \gamma^{k-1} + t$ (or $\gamma^k =
\gamma^{k-1} t$). 
\newline \hspace*{.2in} $\B^k = \Hb \X\Tra\Y + \Hb(\Z^{k-1} - \U^{k-1})$.
\newline \hspace*{.2in} $\Z^k =
\text{SVT}_{\gamma^k}(\B^k+\U^{k-1})$. \hspace{.1in} (Record $\Z^{k}$ at each iteration.)
\newline \hspace*{.2in} $\U^k = \U^{k-1} + \B^k-\U^k$
\newline end
\item[4.]  Output $\{\Z^k:k = 1,\cdots, K\}$ as the algorithmic
  regularization path. 
\end{enumerate}
\end{algorithm}

Then, following the framework of the sparse linear regression example, our
ADMM Algorithmic Regularization Path for the reduced-rank mutli-task
learning (regression) is outlined in Algorithm~\ref{alg_rrr}.  Notice
that the algorithm has the same basic steps as in the previous example
except that solving the proximal operator for the $\Z$ sub-problem
entails singular value thresholding.  This step is the most
computationally time consuming aspect of the algorithm as $K$ total
SVDs must be computed to approximate the sequence of solutions.  Also
note that similarly to the sparse regression example, the inversion
needed, $(\X\Tra\X/n +  \Ib)\Inv$, can be precomputed by using the matrix
inversion identities as previously discussed and cached as a
convenient factorization; hence, this 
is computationally feasible even when $p \gg n$.  

\begin{figure}[!t]
\centering
\includegraphics[width=3.25in]{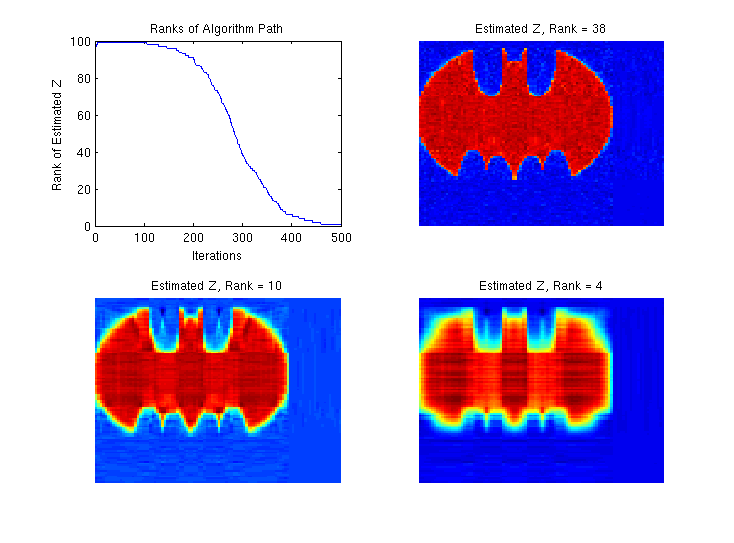}
\caption{Reduced Rank Regression simulated example.  The true
  coefficient matrix, $\B \in \Re^{100 \times 100}$, is an image of
  batman that is rank 38.  Our Algorithm Path provides a sequence of
  low-ranks solutions at a fine resolution (top left) that
  well-approximate the low-rank signal; three such low-rank solutions
  (top right and bottom 
  panel) are shown from iterates of our Algorithm Path. }
\label{fig:rrr}
\end{figure}

\begin{table}[!!h]
\centering
\begin{tabular}{l|r|r|r}
\hline
\hline
& \# Ranks Considered & \# SVDs & Time in Seconds \\
\hline
Algorithm Path & 90 & 476 & 2.354 \\
Proximal Gradient & 57 & 2519 & 12.424 \\
ADMM & 51 & 115,946 & 599.144 \\
\hline
\hline
\end{tabular}
\caption{Algorithm comparisons for reduced rank regression example.}
\label{tab:rrr}
\end{table}

To demonstrate the computational advantages of our approach, we
conduct a small simulation study comparing our method to the two most
commonly used algorithms for reduced-rank regression: proximal
gradient descent and ADMM.  First, we generate data
according to the model: $\Y = \X \B + \mathbf{\epsilon}$, where
$\X_{200 \times 100}$ is generated as independent standard Gaussians,
$\B_{100 \times 100}$ is an image of the Batman symbol, and
$\mathbf{\epsilon}_{200 \times 100}$ is independent standard Gaussian
noise.  We set the signal in the coefficient matrix to be a low-rank
image of the Batman symbol, $\mathrm{rank}(\B) = 38$, which can be
well-approximated by further reduced rank images.  We applied our
Algorithmic Regularization Paths to this simulated example with 500
logarithmically-spaced values of $\gamma$.  Results are
given in Figure~\ref{fig:rrr} and show that our Algorithm Path 
smoothly explores the model space of reduced rank solutions and
nicely approximates the true signal as a low-rank batman
image.  We also conduct a timing comparison to implementations of
proximal 
gradient descent and ADMM algorithms using warm-starts for this same
example; results 
are given in Table~\ref{tab:rrr}.  Here, we see that our approach
requires much fewer SVD computations and is much faster than both
algorithms, especially the ADMM algorithm.  Additionally, both the
ADMM and proximal gradient algorithm employed 100 logarithmically
spaced values of the regularization parameter, $\lambda$.  With this,
however, we see that not all possible ranks of the model space are
considered, 
with proximal gradient and ADMM considering 57 and 51 ranks out of 100
respectively.  In contrast, our Algorithmic Regularization Path yields
a sequence of sparse solutions at a much finer resolution, considering
90 out of the 100 possible ranks.  Thus, for proximal 
gradient and ADMM algorithms to consider the same range of possible
sparsity levels (ranks), a greater number of problems would have to be
solved over a much finer grid of regularization parameters, further
inflating compute times.

Overall, our
approach yields substantial computational savings for computing a
sequence of sparse solutions for reduced rank regression compared to other
state-of-the-art methods for this problem.

\subsection{Convex Clustering}
\label{sec:admm_path_cvx_clustering}

Our final example applies the ADMM Algorithmic Regularization path to
an example with fusion type or non-separable penalties, namely 
a recently introduced convex formulation of cluster analysis
\cite{ChiLan2013,HocVerBac2011,LinOhlLju2011}. Given $n$ points
$\V{y}_1,\ldots,\V{y}_n$ in $\Re^p$, we pose the clustering problem as
follows. Assign to each point $\V{y}_i$ its own cluster center
$\V{\boldsymbol{\beta}}_i \in \Re^p$. We then seek an assignment of
$\V{\boldsymbol{\beta}}_i$ that minimizes the distances between
$\V{y}_i$ and $\V{\boldsymbol{\beta}}_i$ and seeks sparsity between
cluster center pairs $\V{\boldsymbol{\beta}}_i$ and
$\V{\boldsymbol{\beta}}_j$.  Computing all possible cluster
assignments, however, is an NP hard problem.  Hence,  the following
relaxation poses finding the
cluster assignments as a convex optimization problem:
\begin{eqnarray}
\underset{\V{\boldsymbol{\beta}}_1, \ldots, \V{\boldsymbol{\beta}}_n}{\text{minimize}}\; \frac{1}{2}\sum_{i=1}^n \lVert \V{y}_i-\V{\boldsymbol{\beta}}_i \rVert_2^2 + \lambda \sum_{i<j}w_{ij} \lVert \V{\boldsymbol{\beta}}_i-\V{\boldsymbol{\beta}}_j \rVert_2,
\label{eq:objective_function}
\end{eqnarray}
where $\lambda$ is a positive regularization parameter, and $w_{ij}$ is a
nonnegative weight. When $\lambda=0$, the minimum is attained when
$\V{\boldsymbol{\beta}}_i=\V{y}_i$, and each point occupies a unique
cluster. As $\lambda$ increases, the cluster centers begin to
coalesce. Two points $\V{y}_i$ and $\V{y}_j$ with
$\V{\boldsymbol{\beta}}_i = \V{\boldsymbol{\beta}}_j$ are said to
belong to the same cluster. For sufficiently large $\lambda$ all
points coalesce into a single cluster at $\overline{\V{y}}$, the mean
of the $\V{y}_i$. Because the objective in
(\ref{eq:objective_function}) is strictly convex and coercive, it
possesses a unique minimizer for each value of $\lambda$. This is in
stark contrast to other typical criteria used for clustering, which
often rely on greedy algorithms that are prone to get trapped in
suboptimal local minima. Because of its coalescent behavior, the
resulting solution path can be considered a convex relaxation of
hierarchical clustering \cite{HocVerBac2011}.

This problem generalizes the fused LASSO \cite{TibSauRos2005}, and as
with other fused LASSO problems, penalizing affine transformations of
the decision variable makes minimization challenging in general. The one exception is when a 1-norm is used instead of the 2-norm in the fusion penalty terms. In this case, the problem reduces to a weighted
one-dimensional total variation denoising problem. Under other norms, including the 2-norm, the situation, is salvageable if we adopt a splitting strategy discussed
earlier in Section~\ref{sec:review} for dealing with fusion type or non-separable penalties. Briefly, we consider using the 2-norm in the fusion penalty to be most broadly applicable since the solutions to the convex clustering problem
become invariant to rotations in the data. Consequently, clustering assignments will also be guaranteed to be rotationally invariant.

Let the variables $\V{z}_{ij} \in \Re^p$ record the differences
between the $i$th and $j$th points. We denote the collections of
variables $\{\V{\boldsymbol{\beta}}_i\}_{i=1}^n$ and
$\{\V{\boldsymbol{z}}_{ij}\}_{i < j}$ by $\V{\boldsymbol{\beta}}$ and
$\V{z}$ respectively. Then the original problem can be reformulated as: 
\begin{equation}
\label{eq:split_objective_cluster}
\begin{split}
\underset{\V{\boldsymbol{\beta}}, \V{z}}{\text{minimize}}\; \; \frac{1}{2}\sum_{i=1}^n \lVert \V{y}_i-\V{\boldsymbol{\beta}}_i\rVert_2^2 +
\lambda \sum_{i<j} w_{ij} \lVert \V{z}_{ij} \rVert_2 \quad \text{subject to} \quad \V{\boldsymbol{\beta}}_{i} - \V{\boldsymbol{\beta}}_{j} - \V{z}_{ij} = \V{0}.
\end{split}
\end{equation}

Consider the ADMM algorithm derived in \cite{ChiLan2013} for solving
(\ref{eq:split_objective_cluster}). Let $\V{u}_{ij} \in \Re^p$ denote
the Lagrange multiplier for the $ij$th equality constraint. Let
$\V{u}$ denote the collection of variables $\{\V{u}_{ij}\}_{i < j}$. 
The augmented Lagrangian is given by:
\begin{eqnarray*}
\mathcal{L}(\V{\boldsymbol{\beta}},\V{z},\V{u}) = \frac{1}{2} \sum_{i=1}^n \lVert \V{y}_i - \V{\boldsymbol{\beta}}_i \rVert_2^2 + \lambda \sum_{i < j} \VE{w}{ij} \lVert \V{z}_{ij} \rVert_2 
+ \frac{1}{2} \sum_{i < j} \lVert \V{\boldsymbol{\beta}}_i - \V{\boldsymbol{\beta}}_j - \V{z}_{ij} + \V{u}_{ij} \rVert_2^2.
\end{eqnarray*}
Then, the three ADMM subproblems are given by:
\begin{eqnarray*}
\Vn{\boldsymbol{\beta}}{k+1} & = & \underset{\V{\boldsymbol{\beta}}}{\arg\min}\; \frac{1}{2} \sum_{i=1}^n \lVert \V{y}_i - \V{\boldsymbol{\beta}}_i \rVert_2^2 
+ \frac{1}{2} \sum_{i < j} \lVert \V{\boldsymbol{\beta}}_i - \V{\boldsymbol{\beta}}_j - \V{z}_{ij} + \V{u}_{ij} \rVert_2^2 \\
\Vn{z}{k+1} & = & \underset{\V{z}}{\arg\min}\; \lambda \sum_{i < j} \VE{w}{ij} \lVert \V{z}_{ij} \rVert_2 
+ \frac{1}{2} \sum_{i < j} \lVert \V{\boldsymbol{\beta}}_i - \V{\boldsymbol{\beta}}_j - \V{z}_{ij} + \V{u}_{ij} \rVert_2^2 \\
\Vn{u}{k+1}_{ij} &= & \Vn{u}{k}_{ij} + [\Vn{z}{k+1}_{ij} - (\Vn{\boldsymbol{\beta}}{k+1}_i - \Vn{\boldsymbol{\beta}}{k+1}_j)].
\end{eqnarray*}
Splitting the variables in this manner allows us to solve a series of
straightforward subproblems. 
Updating $\V{\boldsymbol{\beta}}$ involves solving a ridge regression
problem. Despite the fact that the quadratic penalty term is not
separable in the $\V{\boldsymbol{\beta}}$, after some algebraic
maneuvering, which is detailed in \cite{ChiLan2013}, it is possible to
explicitly write down the updates for each $\V{\boldsymbol{\beta}}$
separately:
\begin{eqnarray*}
\Vn{\boldsymbol{\beta}}{k+1}_i & = & \left [\frac{1}{1 + n} \V{Y}_i + \frac{n}{1 + n}\overline{\V{y}} \right ] + \frac{1}{1 + n} \left [\sum_{j > i} [\Vn{u}{k}_{ij} + \Vn{z}{k}_{ij} ]-\sum_{j < i}[\Vn{u}{k}_{ji} + \Vn{z}{k}_{ji} ]\right].
\end{eqnarray*}
Updating $\V{z}$ requires minimizing an objective that separates in each of the $\V{z}_{ij}$,
\begin{eqnarray*}
\Vn{z}{k+1}_{ij} & = & \underset{\V{z}_{ij}}{\arg\min}\; \frac{1}{2} \lVert \V{z}_{ij} - [ \Vn{\boldsymbol{\beta}}{k+1}_i - \Vn{\boldsymbol{\beta}}{k+1}_j - \Vn{u}{k}_{ij} ] \rVert_2^2 + \lambda\VE{w}{ij} \lVert \V{z}_{ij} \rVert_2.
\end{eqnarray*}
This step can be computed explicitly using the block-wise
soft-thresholding operator, the proximal operator of the group LASSO
\cite{YuaLin2006}, namely,  
\begin{eqnarray*}
S(\V{z}, \tau) = \underset{\V{\zeta}}{\arg\min}\; \frac{1}{2} \lVert \V{\zeta} - \V{z} \rVert_2^2 + \tau \lVert \V{\zeta} \rVert_2 =  \left [1 - \frac{\tau}{\lVert \V{z} \rVert_2} \right ]_+ \V{z},
\end{eqnarray*}
where $a_+ = \max(a,0)$ and $\tau \geq 0$ controls the amount of
shrinkage towards zero.

For model selection purposes, one typically studies the sequence of
cluster assignments given by coalescent patterns of $\bbeta$, or the sparse
patterns in the first differences of $\bbeta$, as $\lambda$ varies.  We
then seek to quickly approximate this sequence of active sets given by the
coalescent patterns of $\bbeta$ with our Algorithmic Regularization Paths,
summarized in Algorithm~\ref{alg:cluster}.
\begin{algorithm}[t]
\caption{Algorithmic Regularization Path for Convex Clustering}
\begin{enumerate}
\item[1.]  Initialize  $\Vn{z}{0}_{ij} = \V{0}$, $\Vn{u}{0}_{ij} = \V{0}$,
   $\gamma^{(0)} = \varepsilon$, $k = 1$, and set $t>0$.
\item[2.] While $\lVert \Vn{z}{k} \rVert_{\text{F}} > 0$:
\begin{algorithmic}[0]
    		\ForAll{$i$}
		\State
		$\Vn{\boldsymbol{\beta}}{k+1}_i = \left [\frac{1}{1 + n} \V{Y}_i + \frac{n}{1 + n}\overline{\V{y}} \right ] + \frac{1}{1 + n} \left [\sum_{j > i} [\Vn{u}{k}_{ij} + \Vn{z}{k}_{ij} ]-\sum_{j < i}[\Vn{u}{k}_{ji} + \Vn{z}{k}_{ji} ]\right]$
		\EndFor
		\ForAll{$i < j$}
		\State
$\Vn{z}{k+1}_{ij} = S \left(\Vn{\boldsymbol{\beta}}{k+1}_i - \Vn{\boldsymbol{\beta}}{k+1}_j - \Vn{u}{k}_{ij},  \gamma^{(k)}\VE{w}{ij} \right )$
		\State
$\Vn{u}{k+1}_{ij} = \Vn{u}{k}_{ij} + [\Vn{z}{k+1}_{ij} - (\Vn{\boldsymbol{\beta}}{k+1}_i - \Vn{\boldsymbol{\beta}}{k+1}_j)],$
		\EndFor
		\State $\gamma^{(k+1)} = t \gamma^{(k)}$
  \end{algorithmic}
\item[3.] Output $\left \{\Vn{z}{k}_{ij} \right\}$ as the algorithm
  path .
\end{enumerate}
\label{alg:cluster}
\end{algorithm}

As in the general case, we can use iterates of the $\z$-subproblem to
approximate a sparse sequence of cluster assignments.  Given $\V{z}$, we
can determine a clustering assignment in time that is linear in the
number of data points $n$. We simply apply breadth-first search to
identify the connected components of the following graph induced by
the $\V{z}$. The graph identifies a node with every data point and
places an edge between the $i$th and $j$th node if and only if
$\V{z}_{ij} = \V{0}$. Each connected component corresponds to a
cluster.


We now illustrate on a simulated ``halfmoon'' data set of $n = 200$
points in $\Re^2$, that computing our Algorithm Path can lead to non-trivial
computational cost savings for obtaining a sequence of clustering
assingments. We first detail some preliminaries. Although we do not
take 
the space to discuss it here, in practice the choice of weights is
very important. This topic is explored in \cite{ChiLan2013}, and we
use the sparse kernel weights which were shown to work well
empirically in that paper. 
We created a geometric sequence of parameters $\lambda^{(k)}$ and
$\gamma^{(k)}$, namely given a fixed multiplicative factor $t > 1$, we
set $\lambda^{(k+1)} = t \lambda^{(k)}$. The sequence $\gamma^{(k)}$
was constructed similarly, although we study our Algorithm Paths for
several multiplicative factors, $t \in \{1.1,1.05,1.01 \}$. 

In contrast to the regularization path, the Algorithm Path does not require
any convergence checks since only one step is taken at each grid
point. Nonetheless, we only report the number of rounds of ADMM updates
taken by each approach. The Algorithm Path took $259, 1294,$ and
$2,536$ rounds of updates for the three step sizes considered; in
contrast, the regularization path even for a very modest tolerance
level, $10^{-4}$, required a grand total
of 30,008 rounds of updates, substantially more than our approach.

\begin{figure}
\centering
\includegraphics[scale=0.35]{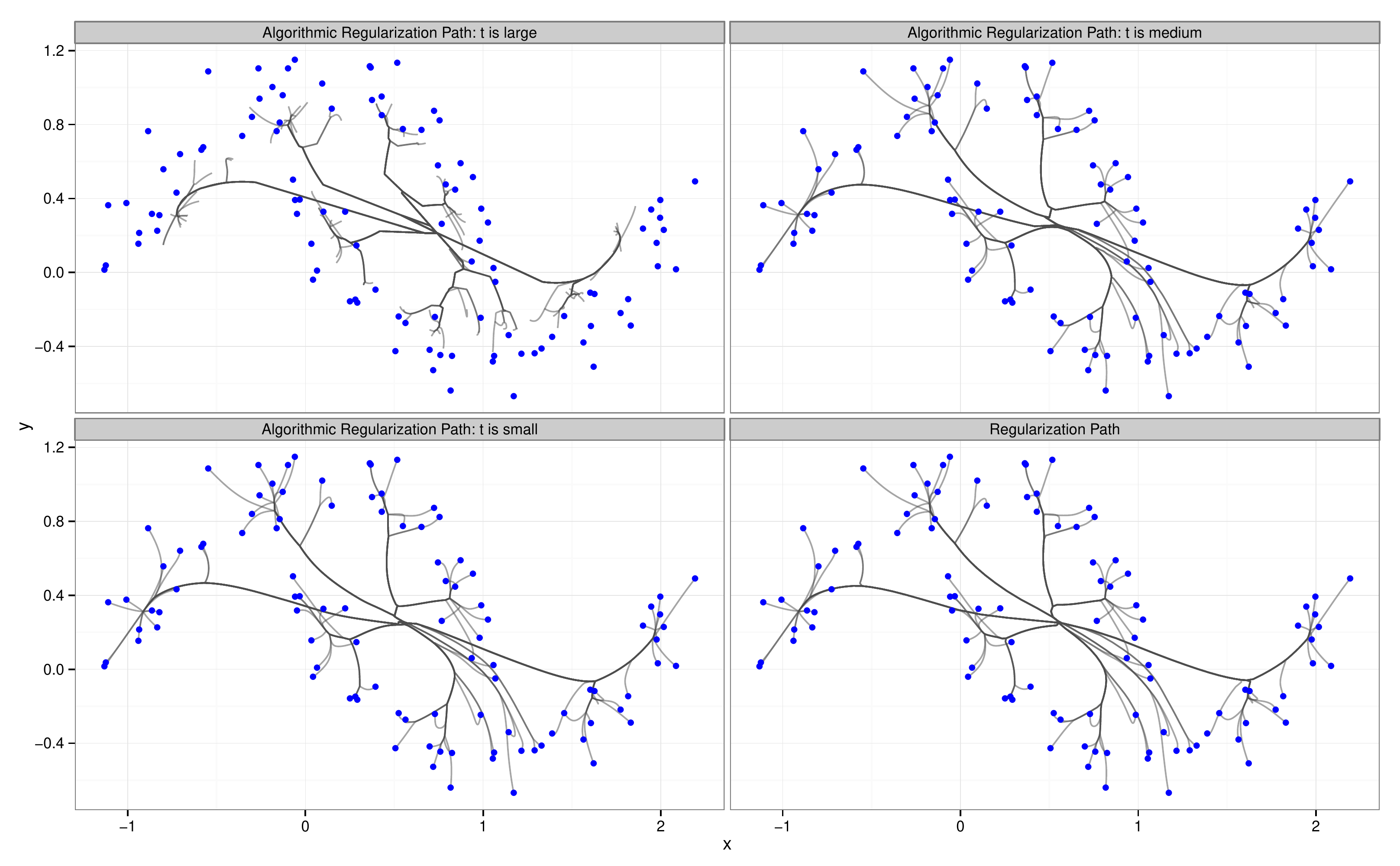}
\caption{Convex clustering on simulated data: In the first three
  panels (from left to right, top to bottom), lines trace the ADMM
  Algorithmic Regularization path of the individual cluster centers as
  the algorithm path parameter $\gamma$ increases for $t = 1.1$
  (large), $1.05$ (medium), and $1.01$ (small). In the panel in the
  lower right corner, the lines trace the regularization path of the
  individual cluster centers as the regularization parameter $\lambda$
  increases. } 
\label{fig:clusterpath_algo_all}
\end{figure}

Figure~\ref{fig:clusterpath_algo_all} shows the ADMM Algorithmic
Regularization paths and regularization path respectively for this
simulated example. For each data point $i$ we plot the sequence of the segments between consecutive estimates of its center, namely $\boldsymbol{\beta}_i^{k+1}$ and $\boldsymbol{\beta}_i^k$. These paths begin to overlap and merge into ``trunks" when center estimates for close-by data points begin to coincide as the parameters $\lambda^{(k)}$ and $\gamma^{(k)}$ becomes sufficiently large.
For sufficiently small step sizes for the regularization levels the
Algorithm Path and regularization path are strikingly similar, as
expected and demonstrated previously in our other examples. For larger
step sizes, however, the paths differ markedly, but still appear to
capture the same clustering assignments.  Overall,
although the simulated data is relatively small, computing the whole
regularization path, even for a modest stopping tolerance can, requires
an order of magnitude more iterations and computational time than the
Algorithm Path.


\section{Discussion}
\label{sec:discussion}

In this chapter, we have presented a novel framework for approximating
the sequence of active sets associated with regularization paths of sparse
statistical learning problems.  Instead of solving optimization
problems over a grid of penalty parameters as in traditional
regularization paths, our algorithm performs a series of one-step
approximations to an ADMM algorithm employing warm-starts with the
goal of estimating a good sequence of sparse models.  Our approach has a 
number of advantages including easy implementation, exploration of the
sparse model space at a fine resolution, and most importantly fast
compute times; we have demonstrated these advantages through
several sparse statistical learning examples.

In our demonstrations, we have focused simply on computing the full
sequence of active sets corresponding to the
regularization path which is the critical computationally intensive
step in the process of model selection.  Once the sequence of sparse
models has been found, common methods for model selection such as AIC,
BIC, cross-validation and stability selection, can be employed to
choose the optimal model.  We note that with regularization paths,
model selection procedures typically choose the optimal $\lambda$
which indexes the optimal sparse model.  For our Algorithm Paths which
do not directly solve regularized statistical problems, model
selection procedures should be used to choose the 
optimal iteration, $k$, and the corresponding sparse model given by
the active set of $\z^{k}$.  While this chapter has focused on finding
the sequence of sparse models via our Algorithm Paths, we plan to
study using these paths in conjunction with common model
selection procedures in future work.

As the ADMM algorithm has been widely used for sparse statistical
learning problems, the mechanics are in place for broad application of
our Algorithm Paths which utilize the three standard ADMM subproblems.
Indeed, our approach could potentially yield substantial computational 
savings for any ADMM application where the $\bbeta$ and $\z$ can be
solved efficiently.  Furthermore, there has
been much recent interest in distributed versions of 
ADMM algorithms~\cite{yin2013parallel, mota2012distributed}.  Thus,
there is the potential to use these in conjunction with our problem to
distribute computation in the $\bbeta$ and $\z$ subproblems and
further speed computations for Big-Data problems.  Also, we have
focused on developing our Algorithm Path for sparse statistical
learning problems that can be written as a composite of a smooth loss
function and a non-smooth, convex penalty.  Our methods, however, can
be easily extended to study constrained statistical learning problems,
such as that of the support vector machines.  Finally, our framework
utilizes the ADMM splitting method, but the 
strategies we develop could also be useful for computing a sequence of
sparse models using other operator splitting algorithms.

Our work raises many questions from statistical and
optimization perspectives.  Further work needs to be done to
characterize and study the mathematical properties of the Algorithm
Paths as well as relate them to existing optimization procedures and
algorithms. For example, ADMM is just one of many variants of proximal methods \cite{parikh2013proximal}. We suspect that other variants, such as proximal gradient descent, used to fit sparse models will also benefit from an Algorithm Path approach in expediting the model selection procedure. We leave this as future work.

In our demonstrations in Section~\ref{sec:examples}, we
suggested empirically that our Algorithm Paths with a tiny step size
closely approximate the sequence of active sets associated with
regularization paths.  Further work needs to be done to verify this
connection mathematically.
Along these lines, a key practical question is how to choose the
appropriate step size for increasing the amount of regularization as
the algorithm progresses.  As we have demonstrated, changing the step size yields paths with very different solutions and behaviors that
warrant further investigation.  For now, our recommendation is to
employ a fairly small step size as these well-approximate the
traditional regularization paths in all of the examples we have
studied.  Additionally, our approach may be related to other new
proposals for computing regularization paths based on partial
differential equations, for example \cite{shinew}; these potential
connections merit further investigation.

Our work also raises a host of interesting statistical
questions as well.  The sparse regression example suggested that
Algorithm Paths may not simply yield computational savings, but may
also perform better in terms of variable selection.  This raises an
interesting statistical prospect that we plan to carefully study in
future work.

To conclude, we have introduced a novel approach to approximating the
sequence of active sets associated with
regularization paths for large-scale sparse statistical learning
procedures.  Our methods yield substantial computational savings and
raise a number of interesting open questions for future research.

\section*{Acknowledgments}

The authors thank Wotao Yin for helpful discussions.  Y.H. and
G.A. acknowledge support from NSF DIMS 1209017, 1264058, and 1317602.
E.C. acknowledges support from CIA Postdoctoral Fellowship \#2012-12062800003.

 \bibliographystyle{spmpsci}
 \bibliography{admm}
\end{document}